# Tuning Magnetotransport in a Compensated Semimetal at the Atomic Scale

Lin Wang[1,2]*, Ignacio Gutiérrez-Lezama[1,2]*, Céline Barreteau[1], Nicolas Ubrig[1,2], Enrico Giannini[1], A.F. Morpurgo[1,2§]

[1] DQMP, Universite de Geneva, 24 quai Ernest Ansermet, CH-1211 Geneva, Switzerland.

[2] GAP, Universite de Geneva, 24 quai Ernest Ansermet, CH-1211 Geneva, Switzerland.

*These authors contributed equally to this work.

§e-mail: Alberto.Morpurgo@unige.ch

**Either in bulk form, or when exfoliated into atomically thin crystals, layered transition metal dichalcogenides are continuously leading to the discovery of new phenomena. The latest example is provided by 1T'-WTe$_2$, a semimetal recently found to exhibit the largest known magnetoresistance in bulk crystals, and predicted to become a two-dimensional topological insulator in strained monolayers. Here, we show that reducing the thickness through facile exfoliation provides an effective experimental knob to tune the electronic properties of WTe$_2$, which allows us to identify the microscopic mechanisms responsible for the observed classical and quantum magnetotransport down to the ultimate atomic scale. We find that the longitudinal resistance and the very unconventional B-dependence of the Hall resistance are reproduced quantitatively in terms of a classical two-band model for crystals as thin as six monolayers, and that for thinner crystals a crossover to an insulating, Anderson-localized state occurs. Besides establishing the origin of the very large magnetoresistance of bulk WTe$_2$, our results represent the first, complete validation of the classical theory for two-band electron-hole transport, and indicate that atomically thin WTe$_2$ layers remain gapless semimetals, from which we conclude that searching for a topological insulating state by straining monolayers is a challenging, but feasible experiment.**

Semimetallic compounds are known to often exhibit many unusual electronic properties, including a non-saturating magnetoresistance (MR) with values among the largest ever reported[1–8]. Different theoretical scenarios –based on charge inhomogeneity[1,3], spin-orbit interaction[5], or the linear dispersion relation of charge carriers[2,5,7,8]– have been proposed in the past to explain such a large MR, but finding experimental evidence to conclusively establish their relevance for experiments has proven extremely difficult. The problem has resurfaced recently with the discovery of a record-high MR in semimetallic[9,10] bulk 1T'-WTe$_2$[6], which was suggested to originate from the classical magnetotransport properties of two, nearly perfectly compensated electron and hole bands. The results of even more recent quantum oscillations[11–14] and angle resolve photoemission (ARPES)[15,16] measurements, however, suggest that more than two bands are present at the Fermi level, and cast doubts about the validity of this conclusion. To address this issue, we have investigated magnetotransport through exfoliated WTe$_2$ crystals to explore the properties of this material from bulk crystals all the way to atomically thin layers.

According to theory[17,18], classical magnetotransport in a two-band nearly compensated semimetal is described by the following expressions for the *B*-dependent longitudinal and transverse resistivity $\rho_{xx}(B)$ and $\rho_{xy}(B)$:

$$\rho_{xx} = \frac{(n\mu_e + p\mu_h) + (n\mu_e\mu_h^2 + p\mu_e^2\mu_h)B^2}{e[(n\mu_e + p\mu_h)^2 + (p-n)^2\mu_e^2\mu_h^2 B^2]} \tag{1}$$

$$\rho_{xy} = \frac{(p\mu_h^2 - n\mu_e^2)B + \mu_e^2\mu_h^2(p-n)B^3}{e[(n\mu_e + p\mu_h)^2 + (p-n)^2\mu_e^2\mu_h^2 B^2]} \tag{2},$$

so that the MR is given by :

$$\frac{\Delta\rho}{\rho} = \frac{\rho_{xx}(B) - \rho_{xx}(0)}{\rho_{xx}(0)}$$
$$= \frac{(n\mu_e + p\mu_h)^2 + \mu_e\mu_h(n\mu_e + p\mu_h)(p\mu_e + n\mu_h)B^2}{(n\mu_e + p\mu_h)^2 + (p-n)^2\mu_e^2\mu_h^2 B^2} - 1 \tag{3}$$

(*n*, *p*, $\mu_e$, and $\mu_h$, are the electron and hole densities and mobility). The regime of near compensation corresponds to having *n* sufficiently close to *p* so that the $B^2$ term in the denominator of Eqs. 1-3 can be neglected in the magnetic field range explored in the experiments (i.e., $(n\mu_e + p\mu_h)^2 \gg (p-n)^2\mu_e^2\mu_h^2 B^2$ for all values of *B* in the measurements). In this case, the MR increases quadratically without saturation, $\frac{\Delta\rho}{\rho} = \mu_e\mu_h B^2$, which is why this scenario was originally invoked to explain the behavior of WTe$_2$[6]. What is peculiar in this regime is the behavior of the Hall resistivity $\rho_{xy}(B)$: it is proportional to *B* at low field and to $B^3$ at high field, with the magnitude and sign of the proportionality coefficients of both terms depending sensitively on the relative values of electron and hole densities (*n* and *p*) and mobility ($\mu_e$ and $\mu_h$). For nearly-compensated systems, small relative changes in *p* and *n* (or in $\mu_e$ and $\mu_h$) can then lead to dramatic changes in the behavior of

$\rho_{xy}(B)$, so that –if the values of the system parameters $n$, $p$, $\mu_e$ and $\mu_h$ can be tuned experimentally– monitoring the evolution of $\rho_{xy}(B)$ allows the validity of the proposed scenario to be proven (or disproven) unambiguously. However, virtually no effort has been devoted so far to investigating $\rho_{xy}(B)$ upon changing the system parameters, largely because of the experimental difficulties involved in controlling and determining unambiguously $p$, $n$, $\mu_e$ and $\mu_h$.

WTe$_2$ is rather unique in this regard because the system parameters can be varied by changing the material thickness through a simple exfoliation process. Additionally, for WTe$_2$, the comparison between experiment and theory is facilitated by the possibility to extract accurate estimates for the parameters directly from the experiments, which drastically narrows down the parameter range when fitting Eqs. 1-3 to the data. We start with the analysis of magnetotransport of bulk crystals. Fig. 1a,b show –as reported in recent earlier studies[6,11–14]– that the relative MR exhibits a large, non-saturating quadratic dependence on $B$, and that Shubnikov-de Haas (SdH) oscillations are clearly visible (for $B \gtrsim 3$ T). The oscillation spectrum exhibits four independent frequencies ranging from approximately 90 to 170 T, whose attribution to different families of charge carriers has not yet been conclusively established[11–14]. We follow Ref. 12 and assume that the Fermi surface is approximately ellipsoidal with a circular cross-section in the plane perpendicular to $B$, which enables us to extract the value of the Fermi momentum $k_F$ and estimate the density of carriers (whose value depends on the assumed degree of anisotropy of the ellipsoidal Fermi surface; see Ref. 12 for details). This estimate serves to fix the starting values of $n$ and $p$ that we use in fitting the data with Eqs. 1-3. From the analysis of the SdH oscillations we also extract fairly precise estimates for $\mu_e$ and $\mu_h$: the value of $B$ at which the SdH oscillations appear fixes the mobility of one type of charge carriers through the condition $\mu B \sim 1$; the mobility for the other carrier type can then be estimated from the magnitude of the quadratic MR, $\frac{\Delta\rho}{\rho} = \mu_e \mu_h B^2$ (determining which type of charge carriers has one or the other mobility value requires the full data analysis).

Starting from the estimated values for of $n$, $p$, $\mu_e$ and $\mu_h$, we perform a fully quantitative fit of the experimental data to Eqs. 1-3. The outcome of the fitting procedure are represented in Fig. 1a and 1c (see also the inset) with red dashed-dotted lines, for both $\rho_{xx}(B)$ and $\rho_{xy}(B)$. We obtain an excellent agreement between measurements and theory with values of parameters that are very close to our initial "rough" estimates. The agreement between theory and data is particularly remarkable and compelling for $\rho_{xy}(B)$: this quantity is "independent" of the measurements used for the initial estimate the model parameters (i.e., $\rho_{xy}(B)$ was not used to estimate the values of the parameters), and yet the analysis fully reproduces at a qualitative and quantitative level the extremely unconventional behavior observed experimentally. As shown in the inset of Fig. 1c, theory even reproduces correctly the change in sign of $\rho_{xy}(B)$ as a function of $B$, whose observation has never been previously reported in a semimetal. As Eq. 2 predicts the zero-crossings to occur at $B^* = \pm\sqrt{\frac{(n\mu_e^2 - p\mu_h^2)}{\mu_e^2\mu_h^2(p-n)}}$, this observation imposes a tight quantitative constraint on the model parameters. It is the unique nature of the behavior observed experimentally, together with the excellent agreement between data and Eqs. 1-3, that allow us to conclude unambiguously that a nearly-compensated two-band model in the classical regime explains the magnetotransport properties of $WTe_2$.

By following the procedure established for macroscopic crystals, we extend our investigations to layers of increasingly smaller thickness, which enables us to explore whether $WTe_2$ behaves as a compensated two-band semimetal even when its thickness is reduced to the atomic scale[19,20]. Transport measurements were performed on devices nano-fabricated on exfoliated flakes whose thickness was determined –all the way to individual monolayers– through a careful analysis relying on Raman spectroscopy[21–23], optical contrast measurements and atomic force microscopy (see Supplementary Information). Fig. 2a shows the MR of exfoliated crystals with thickness down to six monolayers, together with the results (red dashed-dotted lines) of the quantitative fitting to Eqs. 1-3. In all cases, we find that theory fully reproduces all quantitative and qualitative aspects of the data for both $\rho_{xx}(B)$ and $\rho_{xy}(B)$ (see Fig. 2a and 2c). Upon varying the thickness, the longitudinal MR always exhibits a $B^2$ dependence, whereas the functional dependence of $\rho_{xy}(B)$ varies very considerably (presence or absence of non-monotonicity, strength of the non-linearity, concavity/convexity of the curve, etc.). This rich behavior –very precisely reproduced by Eq. 2– is a manifestation of the changes in sign (and magnitude) of the coefficients of the $B$-linear and $B$-cubic terms, determined by the relative magnitude of $n$ and $p$, and of $\mu_e$ and $\mu_h$.

The very systematic agreement between data and Eqs. 1-3 is quite remarkable in two important regards. Firstly, it represents the first complete quantitative validation of the classical theory of transport for a nearly-compensated semi-metal with an electron and a hole band, at a level of detail that has never been reported earlier. Secondly, it shows that a two-band nearly compensated semimetal model does reproduce the magnetotransport properties of WTe$_2$ very satisfactorily in a way that is insensitive to the precise details of the material band structure, which on the energy scale of the band overlap –few tens of meV– is certainly different for the bulk and for crystals that are only six or seven monolayer thick. This insensitivity to details is very likely the reason why a two-band model works so well, despite the presence of more bands, as clearly indicated from SdH oscillations and ARPES measurements. It strongly suggest that electrons (and holes) in different bands have essentially the same mobility, so that classical transport is only sensitive to their total density (and not to the density in each one of the bands). Indeed, if we compare the compensation level *n/p* that we extract from classical transport, with the earlier reported values inferred from the total electron and hole density obtained from the SdH oscillations frequencies[12,13] or ARPES[15,16] measurements in bulk crystals, we find a satisfactory agreement.

As it is apparent from the quality of the agreement between Eqs. 1-3 and the data, the ability to reproduce different qualitative features with a same functional dependence allows all parameters in the model to be extracted precisely. Fig. 2d-f summarize the evolution of *n*, *p*, $\mu_e$ and $\mu_h$ with decreasing thickness, which allows the identification of several trends. Bulk crystals with thickness on the mm scale (i.e., samples B15 and B16) exhibit only small sample-to-sample fluctuations in electron and hole density and mobility: both electron and hole mobility values are rather large (between 5,000 and 10,000 cm$^2$/Vs) and compensation between electrons and holes is nearly perfect (*n/p* ~ 1.1). As the layers are thinned down $\mu_e$ and $\mu_h$ also decrease, because the crystal thickness becomes smaller than the mean free path and scattering at the surface becomes relevant. Nevertheless, even for the thinnest layers analyzed –only six or seven layer thick– $\mu_e$ and $\mu_h$ ~ 1,000 cm$^2$/Vs. The electron and hole densities *n* and *p* exhibit reproducible variations as a function of thickness (the minimum in *n* seen in Fig. 2e, for instance, has been found in the other crystals measured, having thickness between 30 and 50 nm). The origin of these variations is likely the consequence of different physical phenomena, whose relevance depends on the thickness range considered (such as strain for crystals that are several tens of nanometers thick, and the effect of charge transfer from the surface for thinner layers). The net result is that the compensation level worsens for thinner layers and the data show that *n/p* ranges from approximately 0.7 to 1.5 as the thickness is reduced from bulk crystals to crystals that are only seven layer thick. The dependence of $\rho_{xx}(B)$ remains nevertheless quadratic throughout the magnetic field range of our measurements, implying that $(n\mu_e + p\mu_h)^2 \gg (p-n)^2 \mu_e^2 \mu_h^2 B^2$, so that even thin layers still fall in the theoretical regime characteristic of nearly compensated semimetals. The magnitude of the MR is, however, very significantly suppressed, mainly because of the large drop in carrier mobility. We conclude that, although having comparable values for the density of electrons and holes is important, it is the high mobility of the two carriers that is essential to achieve the very large MR measured in WTe$_2$.

As the thickness of WTe$_2$ is decreased even further to approach the ultimate limit of individual monolayers, the transport regime of WTe$_2$ changes qualitatively. The change manifests itself in a metal-insulator-transition clearly visible in the temperature dependence of the conductivity, as shown in Fig. 3a. Interestingly, by plotting the conductivity normalized to the number of layers, we find that the transition occurs when the conductivity-per-layer $\sigma_N \sim e^2/h$. WTe$_2$ crystals that are four layer thick are on the insulating side of the transition –albeit just barely– and the insulating temperature dependence of the conductivity becomes progressively more pronounced for tri and bilayers (we have also fabricated monolayer devices and found that their conductivity is unmeasurably low). In all cases, the observed insulating behavior sets in only at relatively low temperature so that, whereas the room-temperature resistivity increases by only a factor of five when comparing bulk crystals and bilayers, the increase is as large as five orders of magnitude at T = 250 mK.

Identifying the origin of the insulating state is important to fully understand the properties of WTe$_2$ down to the ultimate atomic scale. For crystals only a few monolayers thick, changes in the band structure may reduce the overlap between conduction and valence bands, eventually leading to the opening of a band gap, a scenario that would account for the observed metal-insulator-transition. Such an explanation, however, does not seem consistent with the experimental results. The analysis of magnetotransport, for instance, shows that the density of electrons and holes does not change significantly upon thinning down the material. If anything, the electron density increases, whereas a decrease in band overlap –and the opening of a small gap– should cause the opposite effect. In addition, the conductivity $\sigma$ measured in tri and four layer WTe$_2$ increases steadily with increasing gate voltage $V_G$ (see Fig. 3b), showing no sign of the non-monotonic dependence expected for ambipolar conduction normally observed in the presence of a small band gap. On the contrary, the $\sigma(V_G)$ dependence is consistent with the behavior expected if both types of charge carriers are present, with electrons having a mobility two-to-three times larger than the holes (i.e., the same behavior seen in crystals that are 6-to-12 monolayer thick, see Fig. 2e). In this case, we can estimate $\mu_e$ for trilayers and four layers, by applying the relation $\mu = \frac{1}{C}\frac{d\sigma}{dV_G}$ (with $C$ gate capacitance per unit area) to extend our estimates of carrier mobility; we find $\mu_e \sim 25$ cm$^2$/Vs and $\mu_h \sim 5$ cm$^2$/Vs in the two cases. These values, one-to-two orders of magnitude smaller than those found in the thinnest multilayers that still exhibit metallic behavior (see Fig. 2e), indicate rather unambiguously that the insulating state originates from an increase in disorder strength. Indeed, our observation that the crossover from metallic to insulating behavior occurs when the conductivity-per-layer is $\sigma_N \sim e^2/h$ strongly suggests that carriers in very thin WTe$_2$ layers are Anderson localized.

To find additional experimental evidence supporting the tendency of charge carriers towards localization, we look at magnetotransport measurements of WTe$_2$ crystals that are four layers or thinner. The emergence of quantum corrections to the conductivity in the low temperature MR data –superimposed on the quadratic classical background– becomes clearly apparent as the material thickness is decreased below 10 layers (see Fig. 4a, top panel). The relative effect increases as the layer thickness is decreased. It eventually entirely dominates the MR of crystals thinner than four layers –the same thickness for which the insulating temperature dependence of the conductivity is first observed– which do not any more exhibit the quadratic *B*-dependence expected from Eqs. 1-3. We conclude that, in passing through the metal-insulator-transition, the magnetoresistance of WTe$_2$ changes qualitatively, and its behavior is determined by quantum –and not any more classical– processes. The absolute magnitude of the quantum magnetoconductance at $T = 250$ mK is $\sim e^2/h$ and it decreases upon increasing temperature (see Fig. 3b), as qualitatively expected for the weak antilocalization correction to the conductivity (weak antilocalization occurs because the very high atomic number of tungsten is conducive to a large spin-orbit coupling). Even though the precise nature of the spin-orbit interaction responsible for spin-flip cannot be determined from the measurements, we attempt a semi-quantitative analysis of the data by fitting to Hikami-Larkin-Nagaoka theory, whose expression for the magnetoconductance reads[24]:

$$\Delta\sigma_{xx}(B) = -\frac{e^2}{\pi h}\left[\frac{1}{2}\Psi\left(\frac{1}{2}+\frac{B_\phi}{B}\right) - \frac{1}{2}\ln\left(\frac{B_\phi}{B}\right) - \Psi\left(\frac{1}{2}+\frac{B_\phi+B_{so}}{B}\right) \right.$$
$$\left. + \ln\left(\frac{B_\phi+B_{so}}{B}\right) - \frac{1}{2}\Psi\left(\frac{1}{2}+\frac{B_\phi+2B_{so}}{B}\right) + \frac{1}{2}\ln\left(\frac{B_\phi+2B_{so}}{B}\right)\right] \quad (4)$$

where $\Psi$ is the digamma function, $B_\phi = \frac{h}{8\pi eD}\tau_\phi^{-1}$ is determined by the electron phase coherence time $\tau_\phi$ and $B_{so} = \frac{h}{8\pi eD}\tau_{so}^{-1}$ is determined by the spin relaxation time $\tau_{so}$. Fig. 3b shows that the agreement between measurements performed on trilayer WTe$_2$ and theory is remarkably good. To fit the data we allow $B_\phi$ to vary as a function of temperature (as shown in Fig. 3c), and keep $B_{so}$ constant (= 6 T), as expected from the physical meaning of these parameters ($B_\phi$ is proportional to the inverse of phase coherence time $\tau_\phi^{-1}$, and the approximately linear $T$ dependence of $B_\phi$ agrees with dephasing induced by electron-electron interactions in a diffusive system[25]). Irrespective of the precise values extracted for the parameters, which may depend on the specific theory of weak antilocalization used to fit, the agreement between Eq. 4 and the experimental data, as well as the internal consistency of the behavior observed experimentally, confirm that in WTe$_2$ crystals that are only a few layers thick, localization effects dominate. Our results therefore show that –albeit localized– carriers are still present at the Fermi energy, confirming the results of theoretical calculations[19,20] that indicates how WTe$_2$ remains a gapless semimetal all the way down to monolayer thickness.

The increase in disorder that is responsible for the occurrence of Anderson localization in very thin WTe$_2$ crystals originates from the non-perfect chemical stability of WTe$_2$ in the presence of humidity. Such a non-perfect stability leads to a detectable change in color and contrast (visible under an optical microscope) if thin crystals are exposed to the environment for a sufficiently long time. It appears that between one and two layers on each crystal face are affected by the degradation process during the time needed to nano-fabricate devices (that includes optical inspection to identify the crystals, spinning and backing the resist needed for electron-beam lithography to define the metal contacts, and performing the lift-off process after the metal deposition). As compared to many other exfoliated materials that react chemically when exposed to air –such as phosphorene[26–29] and NbSe$_2$[30], for which strong degradation is visible after one hour of exposure to air– WTe$_2$ seems to be considerably more stable. Significant degradation, giving optically visible signs, is only seen to occur after exposure to the environment for a day, or longer periods. As it has been demonstrated for highly air-sensitive materials like few-layer phosphorene, encapsulation[27–29] (e.g., in between two hBN layers) of WTe$_2$ layers under controlled atmosphere will allow material degradation during the device fabrication process to be eliminated completely. On the basis of the analysis of transport presented in this paper, we anticipate that preserving the chemical integrity of the crystal surface will prevent the decrease in carrier mobility and will enable the realization of atomically thin layers possessing carrier mobility values comparable to those found in the bulk. Such an advance will have important consequences. In particular, it will drastically increase the magnitude of the magnetoresistance in atomic scale crystals of WTe$_2$ to the record values observed in the bulk. It will also make WTe$_2$ fully gate tunable: since mono, bi and trilayers are sufficiently thin to vary their carrier density significantly with electrostatic gate electrodes, the realization of gated encapsulated devices will disclose the possibility to perform experiments that are impossible to perform in thicker, bulk-like crystals. Finally, it will allow mechanical strain in monolayers of $WTe_2$ to be induced and controlled using techniques similar to those employed in graphene[31], thereby enabling experiments to test the recent theoretical prediction that strained monolayers of 1T'-$WTe_2$ are two-dimensional topological insulators[19].

**Methods**

All transport measurements described in this paper have been performed on devices based on $WTe_2$ crystals grown by means of chemical vapor transport using $WCl_6$ as a transport agent, as discussed in more detail in the supplementary information. For measurements on bulk samples, electrical contacts were made with silver epoxy directly on suitably chosen as-grown crystals. To investigate transport on thinner layers, flakes were exfoliated from bulk crystals using adhesive tape and transferred onto a Si substrated covered with 285 nm of $SiO_2$, after which conventional nano-fabrication techniques (electron-beam lithography, metals evaporation, and lift off) were employed to attach electrical contacts (consisting of Ti/Au bilayers, typically 10/70 nm thick). The thickness of the exfoliated crystals was identified by means of atomic force microscopy, optical contrast and Raman spectra as discussed in detail in the supplementary information. Throughout the process of crystal identification and device fabrication, care was taken to minimize exposure of the material to air in order to minimize degradation (the exfoliated crystals where stored in either a glove box with sub-ppm concentratrion of oxygen and water, or in a high vacuum chamber, when not being processed). All magnetotransport measurements were performed either using a Heliox $^3$He system (Oxford instruments) operated in a cryostat with a 14 T magnet and a base temperature of 250 mK, or a cryofree Teslatron cryostat with a 12 T magnet and a base temperature of 1.5 K. The measurements were performed in a current-bias configuration using lock-in amplifiers and home-made low-noise electronic current sources and voltage amplifiers.

**Acknowledgements**

We gratefully acknowledge A. Ferreira for technical support, and M. Gibertini and D. Ki for fruitful discussions. Financial support from the Swiss National Science Foundation and from the EU Graphene Flagship project is also acknowledged.

## Author contributions

L.W. and I.G.L. exfoliated and identified the WTe$_2$ flakes used for the realization of devices. L.W. fabricated the devices and performed the transport measurements. C.B. and E.G. grew and characterized the WTe$_2$ crystals. L.W., I.G.L, and A.F.M. planned the measurements, discussed the results, analyzed the data and wrote the manuscript. L.W. and I.G.L. did Raman measurements together with N.U, to identify the thickness of thin WTe$_2$ exfoliated crystals. All authors carefully read the manuscript and commented on it.

## Additional information

Supplementary information is available. Correspondence and requests for materials should be addressed to A.F.M., L.W. and I.G.L.

## Competing financial interests

The authors declare no competing financial interests.

**Figures**

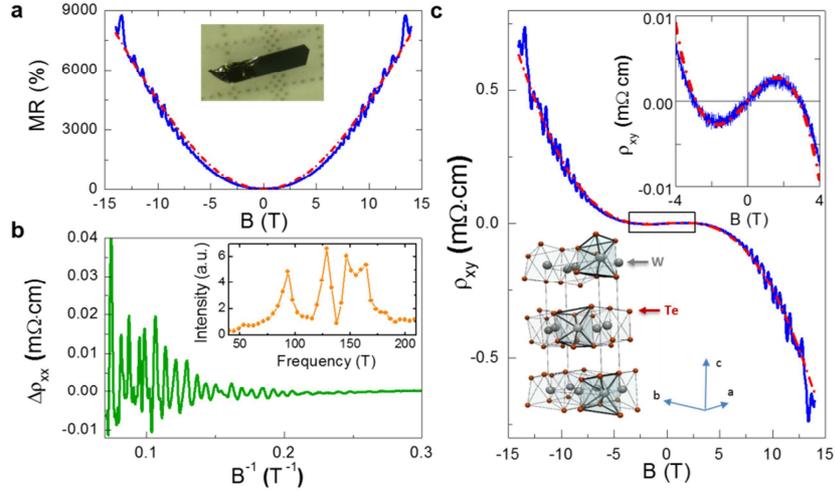

**Figure 1. Magnetotransport of bulk WTe$_2$.** **(a)** Longitudinal magnetoresistance of bulk WTe$_2$ showing a non-saturating, quadratic field dependence –as expected from the classical theory for a compensated semimetal (here and in panel (c) the blue line represents the experimental data and the red dashed-dotted line the theoretical fit with Eqs. 2 and 3, respectively). Shubnikov-de Haas (SdH) oscillations are also clearly visible. The inset shows an optical image of a WTe$_2$ crystal. **(b)** SdH oscillations obtained by subtracting the quadratic background from the longitudinal resistivity, plotted as a function of $B^{-1}$. The dominant peaks in the frequency spectrum, shown in the inset, occur at 93, 129, 147 and 164 T, in virtually perfect agreement with very recent work. **(c)** The Hall resistivity, $\rho_{xy}$, exhibits a very unconventional behavior: it is linear at low $B$ (see inset), proportional to $B^3$ at high fields, and changes sign at approximately $B^* = \pm 3$ T (see inset). This rich behavior –which had never been reported previously for any semimetal– is perfectly captured, at a quantitative level, by Eq. 2 in the main text (represented by the red dashed-dotted line). The inset shows a schematic picture of the crystal structure of WTe$_2$, which is different from that of other common transition metal dichalcogenides with 1T, 2H or 3R structure (it is denoted 1T', because it is similar to a distorted 1T structure). All measurements shown here have been performed at $T = 250$ mK.

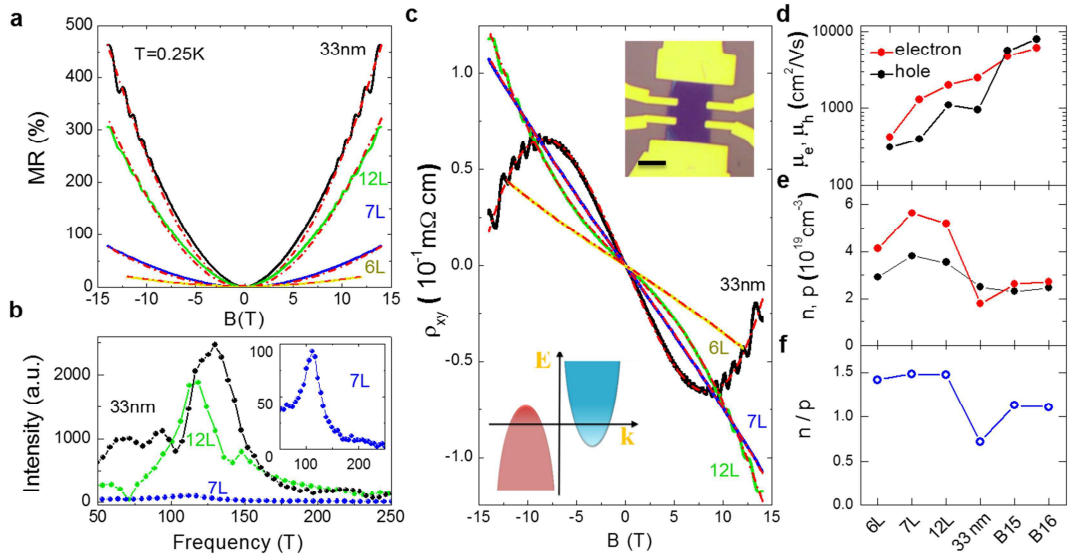

**Figure 2. Evolution of magnetotransport in WTe$_2$ crystals from bulk to the atomic scale.** **(a)** The continuous lines represent the longitudinal MR measured in exfoliated crystals of different thickness (black, 33 nm; green, 12 layers -12L; blue; 7 layers -7L; dark yellow, 6 layers -6L). The red dashed-dotted lines are theoretical fits to Eq. 3 in the main text. Just as for the bulk, the MR does not saturate, is quadratic throughout the experimental range, and is perfectly reproduced by theory. SdH oscillations superimposed on the measurements of all devices are also visible down to a thickness of 7L, starting from increasingly larger magnetic field values. The corresponding spectra of the oscillations are shown in panel **(b)** and its inset; at small thicknesses the smaller number of periods visible in the oscillations decreases the frequency resolution, limiting the visible peak substructure. **(c)** Transverse resistivity $\rho_{xy}$ of the same devices for which the MR is shown in panel (a). The different functional dependencies observed for different thicknesses are fully reproduced by Eq. 2 in the main text (red dashed-dotted lines; all measurements in this figure are done at $T = 250$ mK). The top inset shows an optical image of a 7L device (the scale bar is 5 µm) and the bottom inset a schematic representation of the low-energy electronic structure of a two-band semimetal. Panels **(d,e,f)** summarize the evolution of the electron (red) and hole (black) mobility, density, and their ratio (density), respectively, as extracted from fitting the data to Eqs. 2 and 3.

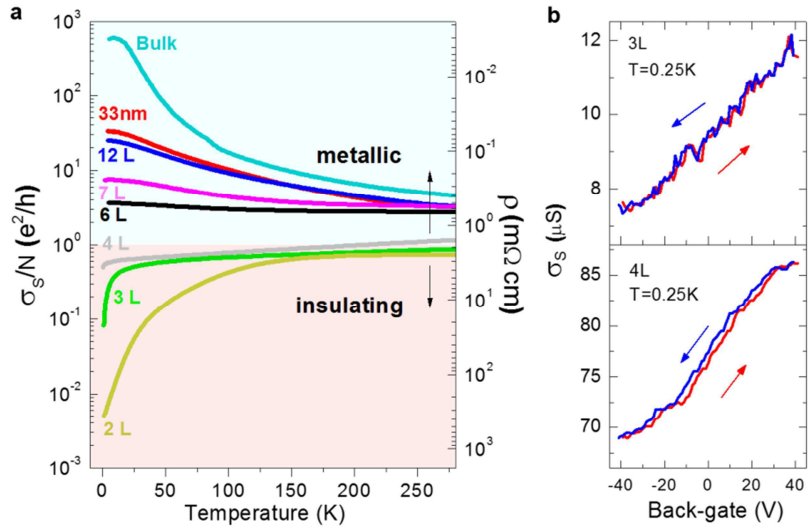

**Figure 3. Metal-insulator transition in atomically thin WTe$_2$. (a)** Temperature dependence of the conductance per square normalized to the number of layers, $\sigma_s/N$, measured in crystals of different thickness (from bulk to bilayer -2L) as a function of temperature $T$ (the right axis gives the corresponding resistivity values). The occurrence of a metal-insulator crossover is clearly apparent, with the six layer (6L) device being still metallic and the four layer one (4L) being the first for which insulating behavior is observed. Note how the transition from metallic to insulator occurs at a conductivity-per-layer that is approximately $e^2/h$. The inset is an optical image of a bilayer device (the scale bar is 5 µm). **(b)** Square conductance of a trilayer (3L, top) and 4L (bottom) as a function of gate voltage, measured at $T = 250$ mK and $B = 0$ T (the crystals are mounted on a highly doped silicon wafer acting as a gate, covered by a 285 nm SiO$_2$ layer acting as gate insulator). The red and blue curves correspond to data taken upon sweeping the gate voltage in opposite directions, as indicated by the arrows of the corresponding color, and illustrate the reproducibility of the measurements.

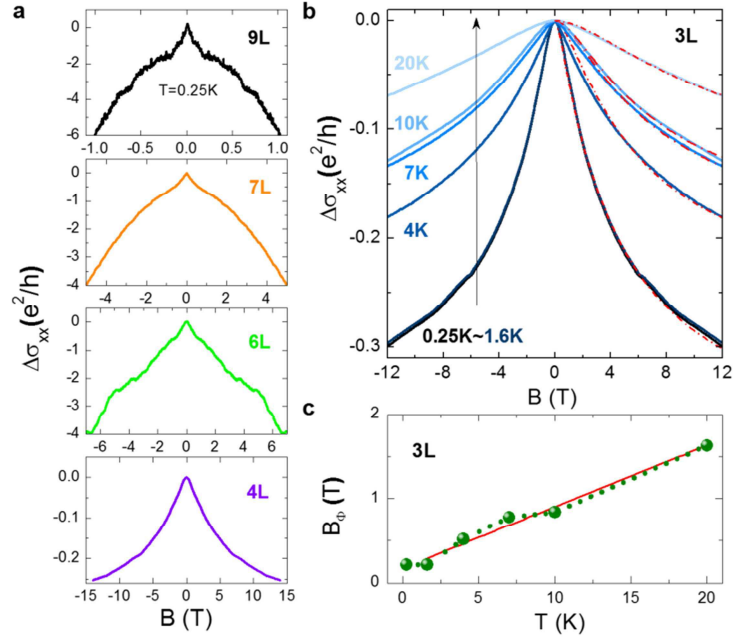

**Figure 4. Quantum localization in atomically thin WTe$_2$ crystals. (a)** Longitudinal magnetoconductance of crystals of different thickness (from top to bottom, 9, 7, 6 and 4 layers, respectively), focusing on the magnetic field range where quantum interference effects are visible (data taken at $T = 250$ mK). Quantum interference manifests itself in the weak antilocalization correction starting to be clearly visible in the 9L device; upon decreasing the crystal thickness the relative magnitude of the effect of quantum interference increases. For the 4L device –the first exhibiting an insulating $T$-dependence of the conductivity– quantum interference dominates magnetotransport, so that no quadratic magnetoresistance of classical origin is visible. **(b)** Magnetic field dependence of the magnetoconductance of a trilayer device (for $T$ ranging 250 mK to 20 K), showing a decrease of the magnetoconductance with increasing temperature, as expected for quantum interference effects. The blue solid lines correspond to the experimental data; the red dashed lines represent theoretical curves obtained by fitting the data with the theory for weak antilocalization, Eq. 4. **(c)** Temperature dependence of $B_\phi$ extracted from fitting the trilayer magnetoconductance with Eq. 4. The linear temperature dependence of $B_\phi$ is consistent with dephasing caused by electron-electron interactions in a diffusive system. The red line is a guide to the eye.

# Supplementary Materials

**1. Growth of WTe$_2$ crystals**

Single crystals of WTe$_2$ were grown by means of chemical vapor transport using WCl$_6$ as a transport agent. As a first step, pure elements W and Te were mixed together with the transport agent WCl$_6$ in a stoichiometric cation ratio, according to the reaction equation:

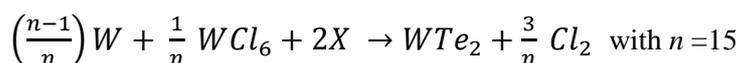

$$\left(\frac{n-1}{n}\right) W + \frac{1}{n} WCl_6 + 2X \rightarrow WTe_2 + \frac{3}{n} Cl_2 \text{ with } n = 15$$

The total material weight used in each growth was about 0.2 - 0.3 g. The mixture was prepared and weighted in a glove box and sealed (under vacuum, $p \sim 5 \times 10^{-6}$ mbar) in a quartz ampule with an internal diameter of 8 mm and a length of 120 mm. The sealed quartz reactor was heated up in a two zone furnace in the presence of a thermal gradient $dT/dx \approx 5 - 10$ °C/cm, with the hot end at $T_{hot} = 890$°C and the cold end at $T_{cold} = 790$°C. After keeping these conditions for four days, the furnace was switched off and cooled down to room temperature. As a result of the process, the whole precursor load moved from the hot side to the cold side of the quartz ampule, where crystals could be found, either isolated or aggregated together. After removing them from the quartz wall, these crystals were either cleaved or cut, and subsequently characterized by X-ray diffraction (XRD) and SEM-EDX analysis. X-Ray diffraction and structure refinement proved that all the crystals crystallized in the orthorhombic $P\ m\ n\ 2_1$ space group[1]. This unique polymorph of WTe$_2$ derives from distortion of the 1T octahedral and is commonly referred to as the "1T' "structure. Within the accuracy of the EDX probe, the atomic ratio W:Te was found to be uniformly equal to 1:2 throughout the crystal inspected, in agreement with the XRD results.

**2. Experimental identification of atomically thin flakes**

WTe$_2$ layers of different thickness –all the way down to individual monolayers– were obtained by means of in-air, micro-mechanical cleaving of bulk crystals with an adhesive tape. The exfoliated crystals were subsequently transferred onto substrates consisting of a highly doped silicon wafer covered with a 285 nm layer of thermally grown SiO$_2$[2], which is known to ensure a good visibility[3,4] of the flakes and can be used as a back gate. More than 30 substrates were systematically inspected under an optical microscope, resulting in the observation of more than 500 atomically thin layers (1 - 9 monolayers thick). An optical picture of one of the many multi-layer flakes found during this work is shown in Fig. S1a. Regions of different intensity are clearly visible, which correspond to layers of different thickness. As it is often the case for two-dimensional materials, including graphene, atomic force microscopy (AFM) measurements of the step height from the substrate does not allow the actual thickness of the exfoliated crystals to be determined, even though measurements performed across neighboring layers do give a step height corresponding to the expected interlayer spacing of $WTe_2$ (0.7 nm, as shown in Fig. S1a). To identify monolayers, we therefore performed a systematic analysis of the optical contrast and Raman spectra of thin $WTe_2$ exfoliated crystals (for crystals thicker than ~ 10 layers, AFM does allow the thickness to be determined with sufficient accuracy for the purposes of the present work).

Our analysis exploits two facts that are known from previous work on atomically thin layers of other transition metal dichalcogenides (TMDs): both the optical contrast of thin exfoliated flakes relative to the substrate [3–5] and their Raman spectra[5–9] evolve systematically with thickness. For this reason, we conducted a combined analysis of the optical contrast, Raman spectra and AFM height profiles of approximately 50 $WTe_2$ exfoliated flakes, whose thickness ranged from 1 to 9 monolayers. A first conclusion of this analysis is that layers exhibiting the same intensity under an optical microscope (i.e., having the same thickness) show reproducibly the same optical contrast[I] and Raman spectra, which makes it possible to univocally assign the layer thickness from a measurement of optical contrast or of Raman spectrum. To assign the thickness correctly, we identified –among the approximately 500 flakes investigated- those parts that have the smallest height relative to the substrate, and found that they systematically correspond to the region of smallest intensity in optical microscope images (even though the actual step height from the substrate is somewhat different in different flakes). Having analyzed a very large number of flakes, we can conclude that these region (exhibiting the smallest intensity values) correspond to individual monolayers. Indeed, since the intensity of monolayers compared to that of the substrate is rather large, thinner layers should have been easily detected if present. Having identified which parts of the exfoliated flakes correspond to monolayers, the thickness of all other layers (bi, tri, etc.) can be easily determined by following the height profile in AFM images, since individual crystalline steps are easily detected in this way.

From this analysis we can then assign the measured optical contrast in the R, G, and B channel[II] and Raman (see discussion below) spectra univocally to the corresponding thickness of atomically thin $WTe_2$ flakes. The result of this assignment is summarized in Fig. S1b (for the RGB contrast), Fig. S2 (for the Raman spectra), and Fig. S3 (for the position and intensity of different Raman peaks that are particularly useful to identify the layer thickness).

---

[I] Note that it is important to use the same microscope and camera –and to keep the same illumination conditions- to have reproducible values of intensity for any given thickness.

[II] The optical contrast is defined as $(I_S - I_F)/I_S$, where $I_S$ and $I_F$ are the intensity of the substrate and the flakes in each of the three RGB channels.

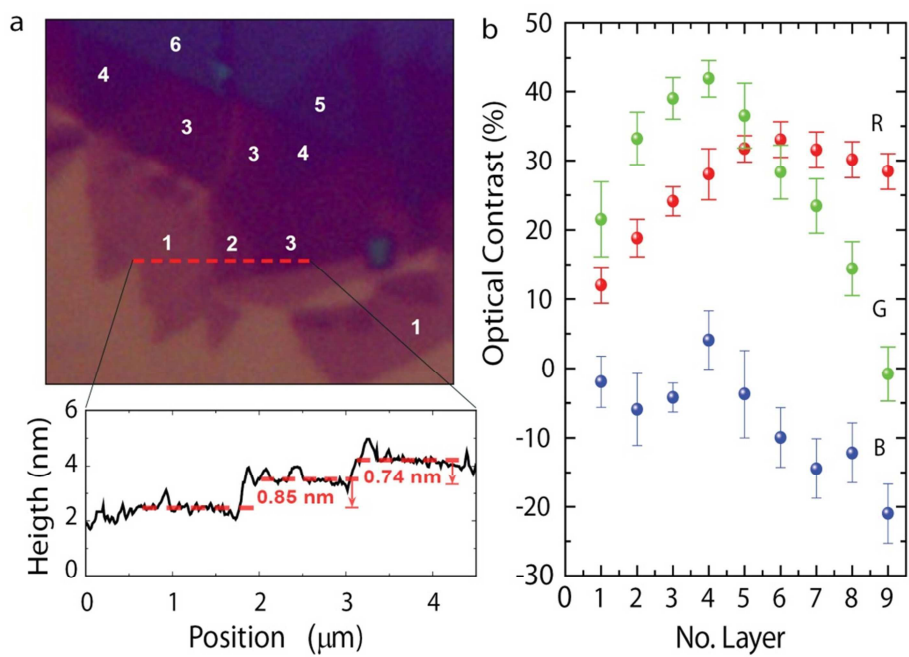

**Figure S1. Atomic force microscopy and relative optical contrast of few-layer WTe$_2$ flakes.** Optical microscope image of an exfoliated WTe$_2$ flake containing layers of different thickness (from 1 to 6 monolayers). The red dash line indicates the position at which the AFM profile shown in the bottom panel was recorded; in the profile, steps corresponding to the expected height of individual WTe$_2$ monolayers are clearly apparent. The attribution of the absolute number of monolayers also requires optical contrast measurements (shown in panel (**b**)) and the analysis of Raman spectroscopy.

The Raman spectra of atomically thin WTe$_2$ crystals, shown in Fig. S2, were measured at room temperature, with a laser wavelength of 514 nm and a power of 0.7 mW. We also measured the spectrum of a bulk crystal, and confined the detailed analysis to flakes that are six layers or thinner, since the spectrum of thicker flakes is too close to that of bulk crystals to allow the unambiguous determination of the layer thickness. With the exception of monolayers, the Raman spectra of all flakes, exhibit seven distinctive sharp peaks in the range from 50 to 250 cm$^{-1}$, whose position is close to that of the peaks measured in bulk crystals: P1 = 80.2, P2 = 90.3, P4 = 117.3, P5 = 132.8, P6 = 164.06 and P7 = 211.6 cm$^{-1}$. Peak P2 is missing in monolayers, which exhibit only six peaks (see Fig. S2), a feature that facilitates their identification (we note that this conclusion could not be made in previous Raman studies of few-layer WTe$_2$[10,11], in which measurements were confined to the range between 100 and 250 cm$^{-1}$). According to a previous theoretical and experimental study of the Raman spectra in WTe$_2$[12] these peaks originate from either A$_1$ (P1, P4 - P7) or A$_2$ (P2 and P3) phonon modes. As for the identification of the layer thickness, the evolution of peaks P1, P3, P4 and P5 is most helpful, as both the position and relative intensities show the largest changes. The analysis of these peaks is summarized in Fig. S3a (peak positions) and Fig. S3b (relative intensity).

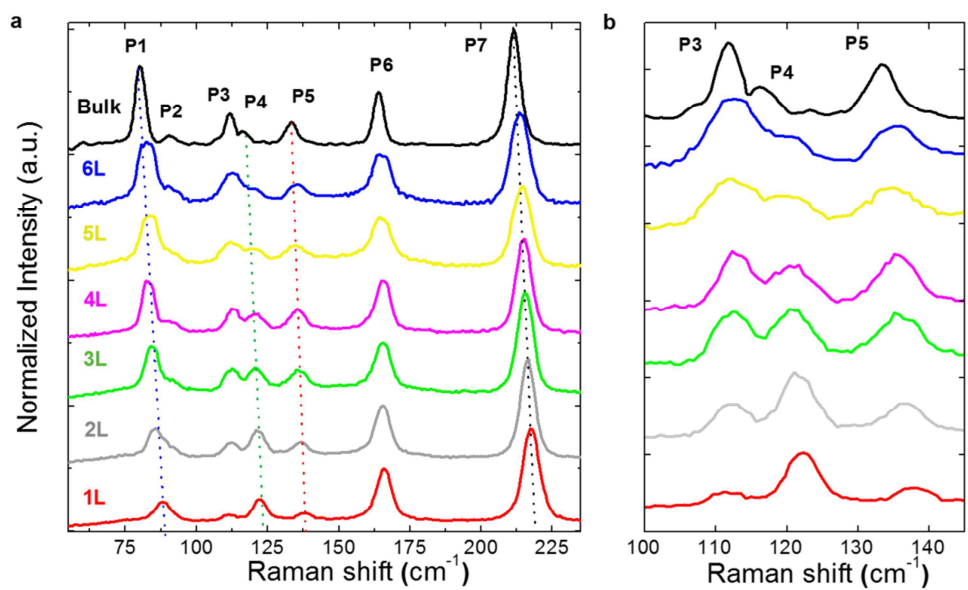

**Figure S2. Raman spectra of bulk and few-layer WTe$_2$. a** Seven peaks, labelled as P1 – P7, are clearly observed in the Raman spectra in the range 50 to 250 cm$^{-1}$ for all thicknesses except for monolayers (peak P2 is missing). The dotted lines are guides to the eye that put in evidence the shift of the peak positions. The range between 100 and 150 cm$^{-1}$, which shows the evolution of peaks P3 – P5, is enlarged in panel **b**. The spectra have been shifted for clarity.

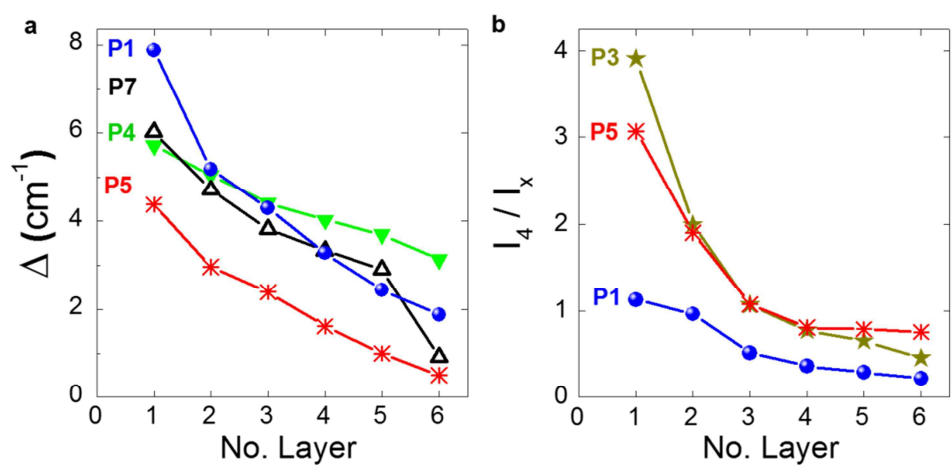

**Figure S3. Evolution of the position and intensity of selected Raman peaks P1 and P3-P5. a** Relative Raman shift Δ of peaks P1, P4, P5 and P7 with respect to their bulk values. **b** Evolution of the ratio between the intensities of peaks P1, P3 and P5 and that of P4.